\documentclass[12pt]{article}

\usepackage{graphicx}

\begin{document}
\begin{center}
{\Large\bf \boldmath  Fundamental Principles of Theoretical Physics and Concepts of Quantum 
Protectorate and Emergence}\footnote{Invited Report, submitted at 
XLVIII All-Russia Conference on Problems in 
Particle Physics, Plasma Physics, Condensed Matter, and Optoelectronics,  
will be held in Moscow 15 - 18 May 2012, Russia.\\
The Conference is dedicated to the 100th anniversary of Professor Yakov Petrovich Terletsky (1912 - 1993), 
the famous theoretical physicist, founder of the Theoretical Physics Department at Peoples' Friendship University of Russia. 
He made an important contribution to Particle and Statistical Physics and the development of higher education in Russia.} \\ 
 
\vspace*{6mm}
{A. L. Kuzemsky   }\\      
{\small \it Bogoliubov Laboratory of Theoretical Physics,}\\
{\small \it  Joint Institute for Nuclear Research, Dubna, Russia.\\
kuzemsky@theor.jinr.ru;   http://theor.jinr.ru/\symbol{126}kuzemsky}         
\end{center}

\vspace*{6mm}

\begin{abstract}
A concise survey of the advanced unifying ideas of modern physics, namely,
spontaneous symmetry breaking, quasiaverages, quantum protectorate and emergence was presented. The interrelation
of the concepts of symmetry breaking, quasiaverages and quantum protectorate was
analyzed in the context of quantum theory and statistical physics.
The main aim of this analysis was to demonstrate the connection and interrelation of these
conceptual advances of the many-body physics and to try to show explicitly that those
concepts, though different in details, have a certain common features. Some problems in
the field of statistical physics of complex materials and systems e.g.  
foundation of the microscopic theory of magnetism   and superconductivity   were pointed in relation to these ideas.
\end{abstract}

\vspace*{6mm}

The development of experimental techniques over the last decades opened the possibility
for studies and investigations of a wide class of extremely complicated and multidisciplinary
problems in physics, astrophysics, biology, material science, etc. In this regard
theoretical physics is a kind of science which forms and elaborates the appropriate
language for treating these problems on the firm ground~\cite{ref1}. This idea was expressed in the
statement of F. Wilczek~\cite{ref2}: "primary goal of fundamental physics is to discover profound
concepts that illuminate our understanding of nature". For example, the theory of
symmetry is a basic tool for understanding and formulating the fundamental notions of
physics. Many fundamental laws of physics in addition to their detailed features possess
various symmetry properties. These symmetry properties lead to certain constraints and
regularities on the possible properties of matter. Thus the principles of symmetry belongs
to the underlying principles of physics. Moreover, the idea of symmetry is a useful and
workable tool for many areas of quantum field theory, physics of elementary particles,
statistical physics and condensed matter physics. Symmetry considerations show that
symmetry arguments are very powerful tool for bringing order into the very complicated
picture of the real world~\cite{ref2,ref3,ref4,ref5,ref6}.\\
It is well known that there are many branches of physics and chemistry where phenomena
occur which cannot be described in the framework of interactions amongst a few
particles. As a rule, these phenomena arise essentially from the cooperative behavior of a
large number of particles. Such many-body problems are of great interest not only
because of the nature of phenomena themselves, but also because of the intrinsic difficulties
in solving problems which involve interactions of many particles ( in terms of known
P.W. Anderson's statement: "more is different"). It is often difficult to formulate a fully
consistent and adequate microscopic theory of complex cooperative phenomena. More
recently it has been possible to make a step forward in solving of these problems. This step
leads to a deeper understanding of the relations between microscopic dynamics and
macroscopic behavior on the basis of emergence concept~\cite{ref6,ref7,ref8,ref9,ref10}.\\
Emergence - macro-level effect from micro-level causes - is an important and profound
interdisciplinary notion of modern science. There has been renewed interest in emergence
within discussions of the behavior of complex systems~\cite{ref6,ref7,ref8,ref9,ref10,ref11,ref12}.
A vast amount of current researches focuses on the search for the organizing principles
responsible for emergent behavior in matter~\cite{ref6,ref7,ref8,ref9,ref10,ref11,ref12}, 
with particular attention to correlated
matter, the study of materials in which unexpectedly new classes of behavior emerge in
response to the strong and competing interactions among their elementary constituents.
As it was formulated by D.Pines~\cite{ref12}: ''we call emergent behavior : the phenomena that
owe their existence to interactions between many subunits, but whose existence cannot be
deduced from a detailed knowledge of those subunits alone''.
There has been renewed interest in emergence within discussions of the behavior of
complex systems and debates over the reconcilability of mental causation, intentionality,
or consciousness with physicalism. This concept is also at the heart of the numerous
discussions on the interrelation of the reductionism and functionalism.\\
In the search for a "theory of everything"~\cite{ref11}, scientists scrutinize ever-smaller
components of the universe. String theory postulates units so minuscule that researchers
would not have the technology to detect them for decades. R.B. Laughlin~\cite{ref8,ref9,ref11}, argued
that smaller is not necessarily better. He proposes turning our attention instead to
emerging properties of large agglomerations of matter. For instance, chaos theory has been
all the rage of late with its speculations about the "butterfly effect," but understanding how
individual streams of air combine to form a turbulent flow is almost impossible~\cite{ref13}. It may
be easier and more efficient, says Laughlin, to study the turbulent flow. Laws and theories
follow from collective behavior, not the other way around, and if one will try to analyze
things too closely, he may not understand how they work on a macro level. In many cases,
the whole exhibits properties that can not be explained by the behavior of its parts. As
Laughlin points out, mankind use computers and internal combustion engines every day,
but scientists do not totally understand why all of their parts work the way they do.
R.B. Laughlin and D. Pines invented an idea of a quantum protectorate~\cite{ref8,ref9,ref11}, "a stable
state of matter, whose generic low-energy properties are determined by a higherorganizing
principle and nothing else". This idea brings into physics the concept that
emphasize the crucial role of low-energy and high-energy scales for treating the
propertied of the substance~\cite{ref8,ref9,ref11,ref12}.
It is known that a many-particle system (e.g. electron gas) in the low-energy limit can be
characterized by a small set of collective (or hydrodynamic) variables and equations of
motion corresponding to these variables. Going beyond the framework of the low-energy
region would require the consideration of plasmon excitations, effects of electron shell
reconstructing, etc. The existence of two scales, low-energy and high-energy, in the
description of physical phenomena was used in physics, explicitly or implicitly.
According to R. Laughlin and D. Pines,''The emergent physical phenomena regulated by
higher organizing principles have a property, namely their insensitivity to microscopics,
that is directly relevant to the broad question of what is knowable in the deepest sense of
the term. The low energy excitation spectrum of a conventional superconductor, for
example, is completely generic and is characterized by a handful of parameters that may be
determined experimentally but cannot, in general, be computed from first principles. An
even more trivial example is the low-energy excitation spectrum of a conventional
crystalline insulator, which consists of transverse and longitudinal sound and nothing else,
regardless of details. It is rather obvious that one does not need to prove the existence of
sound in a solid, for it follows from the existence of elastic moduli at long length scales,
which in turn follows from the spontaneous breaking of translational and rotational
symmetry characteristic of the crystalline state. Conversely, one therefore learns little
about the atomic structure of a crystalline solid by measuring its acoustics. The crystalline
state is the simplest known example of a quantum protectorate, a stable state of matter
whose generic low-energy properties are determined by a higher organizing principle and
nothing else. Other important quantum protectorates include superfluidity in Bose liquids
such as He4 and the newly discovered atomic condensates, superconductivity, band
insulation, ferromagnetism, antiferromagnetism, and the quantum Hall states. The low energy
excited quantum states of these systems are particles in exactly the same sense that
the electron in the vacuum of quantum electrodynamics is a particle: Yet they are not
elementary, and, as in the case of sound, simply do not exist outside the context of the
stable state of matter in which they live. These quantum protectorates, with their
associated emergent behavior, provide us with explicit demonstrations that the underlying
microscopic theory can easily have no measurable consequences whatsoever at low
energies. The nature of the underlying theory is unknowable until one raises the energy
scale sufficiently to escape protection''.\\
The notion of quantum protectorate was introduced to unify some generic features of
complex physical systems on different energy scales, and is a complimentary unifying idea
resembling in a certain sense the symmetry breaking concept, quasiaverages, and so on..
For example, the sources of quantum protection in high-$T_c$ superconductivity and low-dimensional
systems were discussed as well in their study. According to P.W. Anderson
and D.Pines, the source of quantum protection is likely to be a collective state of the
quantum field, in which the individual particles are sufficiently tightly coupled that
elementary excitations no longer involve just a few particles, but are collective excitations
of the whole system. As a result, macroscopic behavior is mostly determined by overall
conservation laws.\\
In our interdisciplinary review~\cite{ref6} we analyzed the applications of the symmetry
principles to quantum and statistical physics in connection with some other branches of
science. The profound and innovative idea of quasiaverages formulated by
N.N.Bogoliubov, gives the so-called macro-objectivation of the degeneracy in domain of
quantum statistical mechanics, quantum field theory and in the quantum physics in
general. We discussed also the complementary unifying ideas of modern physics, namely:
spontaneous symmetry breaking, quantum protectorate and emergence. The interrelation
of the concepts of symmetry breaking, quasiaverages and quantum protectorate was
analyzed in the context of quantum theory and statistical physics.
The main aim of that paper were to demonstrate the connection and interrelation of these
conceptual advances of the many-body physics and to try to show explicitly that those
concepts, though different in details, have a certain common features. Many problems in
the field of statistical physics of complex materials and systems (e.g. the chirality of
molecules) and the foundation of the microscopic theory of magnetism [14,15] and
superconductivity~\cite{ref15} were discussed in relation to these ideas.
It is worth to emphasize once again that the notion of quantum protectorate complements
the concepts of broken symmetry and quasiaverages by making emphasis on the hierarchy
of the energy scales of many-particle systems. In an indirect way these aspects of
hierarchical structure arose already when considering the scale invariance and
spontaneous symmetry breaking in many problems of classical and quantum physics.\\
It was shown also in papers~\cite{ref6,ref15,ref16}  that the concepts of symmetry breaking perturbations
and quasiaverages play an important role in the theory of irreversible processes as well.
The method of the construction of the nonequilibrium statistical operator~\cite{ref16} becomes
especially deep and transparent when it is applied in the framework of the Bogoliubov's
quasiaverage concept. For detailed discussion of the Bogoliubov's ideas and methods in the
fields of nonlinear oscillations and nonequilibrium statistical mechanics see Refs. [15,16].
Thus, it was demonstrated in Ref.~\cite{ref6} that the connection and interrelation of the
conceptual advances of the many-body physics discussed above show that those concepts,
though different in details, have complementary character.\\
To summarize, the ideas of symmetry breaking, quasiaverages, emergence and quantum
protectorate play constructive unifying role in modern theoretical physics. The main
suggestion is that the emphasis of symmetry breaking concept is on the symmetry itself,
whereas the method of quasiaverages emphasizes the degeneracy of a system. The idea of
quantum protectorate reveals the essential difference in the behaviour of the complex
many-body systems at the low-energy and high-energy scales. Thus the role of symmetry
(and the breaking of symmetries) in combination with the degeneracy of the system was
reanalyzed and essentially clarified within the framework of the method of quasiaverages.
The complementary notion of quantum protectorate might provide distinctive signatures
and good criteria for a hierarchy of energy scales and the appropriate emergent behavior.
It was demonstrated also that the Bogoliubov's method of quasiaverages plays a
fundamental role in equilibrium and nonequilibrium statistical mechanics and quantum
field theory and is one of the pillars of modern physics.\\ 
We believe that all these concepts
will serve for the future development of physics~\cite{ref17} as useful practical tools. Additional
material and discussion of these problems can be found in recent publications~\cite{ref18,ref19}.
\end{document}